\documentclass[usenatbib]{mn2e}
\usepackage{amsmath}
\usepackage{graphicx}
\usepackage{natbib}
\def\apj{{ApJ}}                 
\def\mnras{{MNRAS}}             

\usepackage{epstopdf}
\usepackage{color}

\title[effects of correlation on halo formation]{
Effects of correlation between merging steps on the global halo 
formation}

\begin{document}
\author[Pan et al.]{
Jun Pan$^{1}$\thanks{jpan@pmo.ac.cn}, 
Yougang Wang$^{2}$, 
Xuelei Chen$^{2}$,
and Lu\'{i}s Teodoro$^{3}$\\
$^1$ The Purple Mountain Observatory, 2 West Beijing Road,
Nanjing 210008, China\\
$^2$ National Astronomical Observatories, Chinese Academy of Sciences,
Beijing 100012, China\\
$^3$ Department of Physics and Astronomy, University of Glasgow, 
Glasgow G12 8QQ, UK
}
\maketitle

\begin{abstract}
The excursion set theory of halo formation is modified by 
adopting the fractional Brownian motion, 
to account for possible correlation between merging steps. 
We worked out analytically the conditional 
mass function, halo merging rate and formation time distribution 
in the spherical collapse model. We also developed an approximation 
for the ellipsoidal collapse model and applied it to the calculation of 
the conditional mass function and the halo formation time distribution.
For models in which the steps are positively 
correlated, the halo merger rate is enhanced when the accreted 
mass is less than $\sim 25M^*$, while for the negatively correlated 
case this rate is reduced. Compared with the
standard model in which the steps are uncorrelated,
the models with positively correlated steps produce more 
aged population in small mass halos and more younger population 
in large mass halos, while for the models with negatively correlated steps
the opposite is true. An examination of simulation results shows that 
a weakly positive correlation between successive merging steps
appears to fit best. We have also found a systematic
effect in the measured mass function due to the finite volume of simulations.
In future work, this will be included in the halo model to 
accurately predict the three
point correlation function estimated from simulations.

\end{abstract}


\section{introduction}

The excursion set theory provides a simple and intuitive model for 
cosmic structure formation. In this theory, as one varies the smoothing
scale $R$, the linear density fluctuation $\delta_R$ obtained with 
smoothing window $W(R)$ forms a one dimensional random walk. 
It is deemed that the non-linear halo formation and evolution 
history can be treated with the corresponding excursion set theory: 
a halo is formed when a pre-set barrier $\delta_c(z)$ is exceeded by 
a tracjectory of the random walk. 

It is convinient to take the variance of the random density field 
$S(R)=\sigma^2(R)$ as the pseudo-time variable. 
Properties of the random walk rely on the window function 
used and the nature of 
the primordial fluctuation. For Gaussian fluctuations
and using the sharp k-space filter, the random walk produced by 
smoothing is a normal Brownian motion 
(hereafter NBM), i.e. there is no correlation between steps.
The density of trajectories $Q(S,\delta)$ passing through $(S,\delta)$
then satisfies a diffusion equation 
\begin{equation}
\frac{\partial Q}{\partial S}=\frac{1}{2} \frac{\partial^2 Q}{\partial \delta^2}
\end{equation}
From $Q(S,\delta)$ one could derive the halo mass function and merger
rates, etc. \citep{BondEtal1991,LaceyCole1993}.

However, although the N-body simulation results generally agree with
the predictions of the excursion set theory, there are significant 
deviations in the details. The discrepancy is particularly severe in 
the description of small mass halos. This discrepancy
can be partly overcomed by replacing the spherical collapse 
model with ellisoidal collapse model, i.e. by
introducing a moving barrier instead of a fixed barrier imposed on
the random walk  \citep[e.g.][]{ShethTormen2002}.
This practical approach provides a reasonably good fitting formula 
for halo mass function \citep{WarrenEtal2006}, but its prediction 
on the formation time distribution of low- and intemediate-mass
 halo remains unsatisfactory \citep{GiocoliEtal2007}.
One suspects that if correlations between steps of the
random walk is introduced, the excursion set theory might be improved.
Indeed, if the smoothing window function is not a sharp k-space 
filter but a Gaussian filter or a real space tophat filter, the excursion 
steps would be correlated.

In a previous work by the leading author, 
the fractional Brownian motion (FBM), the simplest random walk with
steps correlated in long range, was introduced to generalize 
the excursion set theory. The correlation between steps of the 
random walk was shown to be capable of modifying the final halo 
mass function in a non-trivial way \citep{Pan2007}.
The model presented there was incomplete though, as the solution 
for ellipsoidal collapses and the treatment of halo merging history were 
not discussed. These are the topics of the present report.

This paper is organized as follows. In section 2 we fit the FBM into
the excursion set theory to account for the possible correlated halo formation
process. Then in section 3  the diffusion equation of FBM is solved in 
the spherical collapse model, and we calculate the conditional mass function, 
halo merger rate and the halo formation time distribution with the 
modified theory. Treatment to 
the ellipsoidal collapse is given
in section 4, together with a comparison between theoretical predictions
and measurements with simulations. The final section contains
our conclusions and discussion. We adopt the flat $\Lambda$CDM model with 
the following set of cosmological parameter values: $\Omega_m=0.3$, 
$\Omega_\Lambda=0.7$, $h=0.7$ and $\sigma_8=0.9$.

\section{Modeling correlated merging steps}
\subsection{The diffusion equation}

To work out the halo conditional mass function, the central
element is the diffusion equation which governs the behavior of 
the random walk. Such diffusion equation in turn depends on 
the understanding of the random walk with which the physical 
problem is concerned with. The conditional mass function is 
in fact a two-barriers crossing problem of a random walk, which means
the key object we shall check is the scaling relation between 
$\delta(S_1)-\delta(S_0)$ and $S_1-S_0$.

The FBM is a random process $X(t)$ on
some probability space such that: 
\begin{enumerate}
\item $X(t)$ is continuous and $X(0)\equiv 0$; 
\item for any $t\ge 0$ and $\tau> 0$, the increment $X(t+\tau)-X(t)$ follows 
a normal distribution with mean
zero and variance $\tau^{2\alpha}$, so that
\begin{equation}
P\left( X(t+\tau)-X(t)\le x \right)=\frac{\tau^{-\alpha}}{\sqrt{2\pi}}
\int_{-\infty}^{x} e^{-u^2/2\tau^{2\alpha} }du \ .
\end{equation}
\end{enumerate}
The parameter $\alpha$ is the Hurst exponent, if $\alpha=1/2$, it is reduced to 
the normal Brownian motion \citep[c.f.][]{Feder1988}.
It is also easy to see that the following is satisfied by the FBM:
\begin{equation}
\begin{aligned}
\langle \left[ X(t+\tau)-X(t) \right]^2 \rangle &=\tau^{2\alpha}\\
\langle X(t) \left[X(t+\tau)-X(t)\right] \rangle &=\frac{ (
t+\tau)^{2\alpha}-t^{2\alpha}-\tau^{2\alpha}}{2} ,
\end{aligned}
\label{eq:cov}
\end{equation}
With this definition of FBM, the trajectory density
$Q_\alpha(X, t)$ at time $t$ in interval $(X, X+dX)$ 
follows the diffusion equation \citep[c.f.][]{Lutz2001},
\begin{equation}
\frac{\partial Q_\alpha}{\partial t}=
{\mathcal D}\frac{\partial^2Q_\alpha}{{\partial X}^2}\ ,\  
{\mathcal D}=\frac{1}{2}\frac{d}{dt}\langle X(t)^2\rangle=\alpha t^{2\alpha-1}.
\label{eq:diff}
\end{equation}

Why invoke FBM? 
The trajectory of $\delta(S)$ is 
characterized by properties of the increments 
$\delta(S_1)-\delta(S_0)$ between any pairs of $(S_0, S_1>S_0)$.
If the smoothing window function corresponding to halo definition is not 
a sharp k-space filter but e.g. a Gaussian or top-hat, it is 
easy to check that $\langle [\delta(S_1)-\delta(S_0)]^2 \rangle$ is 
not $ S_1-S_0$ but proportional to $(S_1-S_0)^\alpha$ with a constant 
$\alpha$ in broad scale range. 
The scaling relation is the flagging attribute of the FBM\footnote{There 
are many anomalous random walks such as the fractal time 
process \citep{Lutz2001} which have the same scaling feature and 
are classified as sub-diffusion. The FBM is the simplest one of them.}, 
so it is reasonable to install the FBM into the excursion set theory 
as the simplest approach to analytically inspect halo 
models constructed with random walks with correlated steps.

On the other hand, we know in practice that the boundary, and
subsequently the mass of a halo identified in a simulation is 
rather arbitrary, e.g. the mass picked up is often defined 
by a halo's virial radius while the mass outside is simply 
not counted. The $S$ from the virial mass $M$ is very 
likely not the real place where the particular random walk corresponding 
to the halo hit the barrier. 
In fact \citet{CuestaEtal2007} argued that if one replaces the
virial mass with the static mass, the mass function agrees with the 
Press-Schechter formula remarkably well, rather than the Sheth-Tormen 
one, and the ratio of the virial mass to 
the static mass depends on redshift and halo mass. 

While the true scale can be smaller or larger than the $S$ inferred 
from $R(M)$, one can always parametrize the true scale with  $S(R)$ so that 
$\langle \delta^2 \rangle = S^{2\alpha}$. Taking the approximation that 
$\alpha$ is a constant not too far away from $1/2$, we can 
comfortably assume the
applicability of the FBM.

For an FBM with $\langle \delta^2 \rangle = S^{2\alpha}$, by Eq.~(\ref{eq:cov}) 
,  the variance of the increments at two points 
$(S_0, S_1>S_0>0)$ satisfies
\begin{equation}
\begin{aligned}
\langle \left[ \delta(S_1)-\delta(S_0)\right]^2\rangle = & (S_1-S_0)^{2\alpha} \ ,\\
& {\rm with} \ \alpha \in (0, 1)\ , \ S_1>S_0>0\ ,
\end{aligned}
\end{equation}
so that a new trajectory is formed by 
$\widetilde{\delta}(\Delta S)=\delta(S_1)-\delta(S_0)$
along $\Delta S=S_1-S_0$, given any source 
point $(S_0, \omega_0=\delta_0(S_0))$.
Apparently the new random walk also
complies with the definition of FBM, its diffusion 
equation being
\begin{equation}
\frac{\partial Q_\alpha}{\partial \Delta S}=
{\mathcal D}_{\Delta S} \frac{\partial^2 Q_\alpha}{ {\partial \widetilde{\delta}}^2}\ , \quad
{\mathcal D}_{\Delta S}=\frac{1}{2}\frac{d\langle {\widetilde{\delta}}^2\rangle}{d\Delta S}
=\alpha \Delta S^{2\alpha-1}\ .
\label{eq:dfFBM}
\end{equation}
It is this diffusion equation that ought to be solved to figure out the
conditional mass function, which can be further transformed to the familiar 
diffusion equation of normal Brownian motion
\begin{equation}
\frac{\partial Q_\alpha }{ \partial \widetilde{S} }= \frac{1}{2} 
\frac{\partial^2 Q_\alpha }{{\partial \widetilde{\delta}}^2}
\label{eq:dfNBM}
\end{equation}
by the substitution $\widetilde{S}=\Delta S^{2\alpha}$.

\subsection{The correlation}
Imagine a halo is formed by the collapsing condition $\delta(S_1)=\omega_1$
at $S_1=\sigma^2(M_1)$ with $M_1$ being the halo mass. After some time the halo
merges into a bigger halo with mass $M_0>M_1$ at $S_0=\sigma^2(M_0)<S_1$ 
by another collapsing condition $\delta(S_0)=\omega_0 < w_1$.  The following 
question needs to be addressed:
how the formation event by $\omega_0$ at $S_0$ is correlated with
the past merging process of $\omega_1-\omega_0$ within $S_1-S_0$?

If this process is approximated by the FBM, the correlation function 
can be easily calculated with Eq.~(\ref{eq:cov}). We have
\begin{equation}
\begin{aligned}
\xi_{0,0\rightarrow 1}&=\langle (\omega_0-0)(\omega_1-\omega_0) \rangle\\ &
=\frac{S_1^{2\alpha}}{2}
\left[ 1-\left( \frac{S_0}{S_1} \right)^{2\alpha}-\left( 1-\frac{S_0}{S_1}\right)^{2\alpha}
\right] \ .
\end{aligned}
\end{equation}

\begin{figure}
\resizebox{\hsize}{!}{\includegraphics{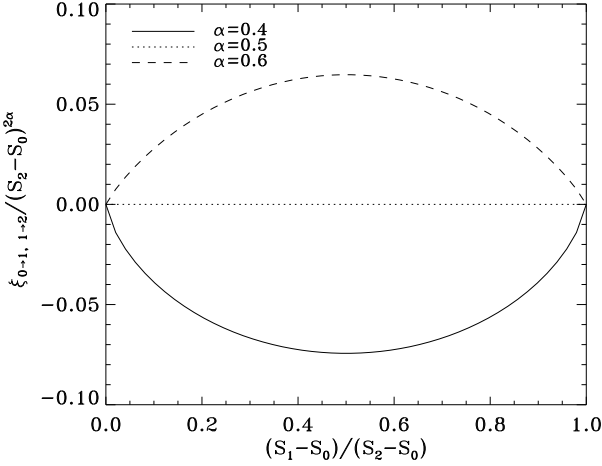}}
\caption{Correlation functions between successive halo merging steps 
(Eq.\ref{eq:corr}).}
\label{fig:corr}
\end{figure}

It is more instructive to capture the correlation between two successive merging steps:
$\omega_2-\omega_1$ within $S_2-S_1$, and $\omega_1-\omega_0$ 
within $S_1-S_0$ ($S_2>S_1>S_0$, $\omega_2>\omega_1>\omega_0$).
The correlation function is similarly
\begin{equation}
\begin{aligned}
\xi_{0\rightarrow1, 1\rightarrow 2}& =  \langle (\omega_2-\omega_1)(\omega_1-\omega_0)
\rangle = \frac{(S_2-S_0)^{2\alpha}}{2} \\ 
\times & \left[ 1- 
\left( \frac{S_1-S_0}{S_2-S_0}\right)^{2\alpha} -
\left(1-\frac{S_1-S_0}{S_2-S_0}\right)^{2\alpha}\right]\ ,
\end{aligned}
\label{eq:corr}
\end{equation}
which is positive when $\alpha>1/2$ and negative if $\alpha<1/2$. 
Apparently, $\alpha=1/2$ causes null correlation (Fig.~\ref{fig:corr}). 
Thus a unified paradigm is 
given by a single parameter controlled process: 
an anti-persistent FBM predicts that a merging step is anti-correlated 
with its immediate early occurrence of merging, and persistent FBM models 
positive correlation.

\section{Spherical collapse}

\subsection{Conditional mass function}
It is quite straightforward to solve the diffusion equation
in spherical collapse model, we can actually copy 
the solution in the literatures 
\citep[e.g.][]{BondEtal1991, LaceyCole1993}. 
Spherical collapse is equivalent to employ the boundary condition that 
there is a fixed absorbing barrier of height 
$\widetilde{\delta}_c=\omega_1-\omega_0 > 0$ to the reformed random walk
described by Eq.~(\ref{eq:dfNBM}).
The number of trajectories of the reformed walk within 
$(\widetilde{\delta}, \widetilde{\delta}+d\widetilde{\delta})$ at 
$\widetilde{S}$ is
\begin{equation}
Q_\alpha d\widetilde{\delta}=
\frac{1}{\sqrt{2\pi \widetilde{S}}} \left[ e^{-\widetilde{\delta}^2/2\widetilde{S}}-
e^{-(\widetilde{\delta}-2\widetilde{\delta}_c)^2/2\widetilde{S}} \right] 
d\widetilde{\delta}\ ,
\end{equation}
and the number of trajectories absorbed by barrier within 
$(\widetilde{S}, \widetilde{S}+d\widetilde{S})$ is given by
\begin{equation}
\begin{aligned}
f(\widetilde{S}, \widetilde{\delta}_c)d\widetilde{S}= &- d\widetilde{S} 
\frac{\partial}{\partial \widetilde{S} } 
\int_{-\infty}^{\widetilde{\delta}_c} Q_\alpha d \widetilde{\delta}  \\
= &\frac{\widetilde{\delta}_c}{\sqrt{2\pi}} \widetilde{S}^{-3/2} \exp\left( 
\frac{-\widetilde{\delta}_c^2}{2\widetilde{S}} \right) d\widetilde{S}\ ,
\end{aligned}
\end{equation}
which directly yields the universal conditional halo mass function
\begin{equation}
\begin{aligned}
f(S_1 & -S_0 ,  \omega_1-\omega_0)dS_1=f(S_1, \omega_1| S_0, \omega_0)dS_1 \\
= &\frac{2\alpha}{\sqrt{2\pi}} 
\frac{\omega_1-\omega_0}{(S_1-S_0)^{\alpha+1}}
\exp\left( -\frac{(\omega_1-\omega_0)^2}{2(S_1-S_0)^{2\alpha}} \right)
dS_1\ .
\end{aligned}
\label{eq:sccmf}
\end{equation}
If $\alpha=1/2$ we recover the Eq.~(2.15) in \citet{LaceyCole1993}.

The conditional mass function is in fact 
more fundamental than the mass function,
since the mass function can be recovered from the conditional mass function
by setting the limit $S_0\rightarrow 0, \omega_0 \rightarrow 0$ \citep{Pan2007},
\begin{equation}
f(S, \delta_c)dS=\frac{2\alpha}{\sqrt{2\pi}} \frac{\delta_c}{S^{\alpha+1}}
\exp\left( {-\frac{\delta_c^2}{2 S^{2\alpha}} } \right)dS\ .
\label{eq:scmf}
\end{equation}

Thus, prompted by the relation between mass function and conditional 
mass function, we recognize that there is  a systematical effect due
to the finite volume of simulations. This can lead to an 
underestimate of the halo mass function in the large mass regime 
(more details are discussed in Appendix A).
Recall that the halo model tends to over-estimate the amplitude 
of the three point correlation function of dark matter, which can 
be corrected (at least partly) by 
applying certain arbitrary high mass cut-off to the halo mass 
function \citep{WangEtal2004, FosalbaPanSzapudi2005}.
To predict the three point correlation function of
a simulation, a better approach would be to use the halo mass function
of Eq.~(\ref{eq:simmf}) to include the finite volume effects.

\subsection{Merger rate}

\begin{figure*}
\resizebox{\hsize}{!}{\includegraphics{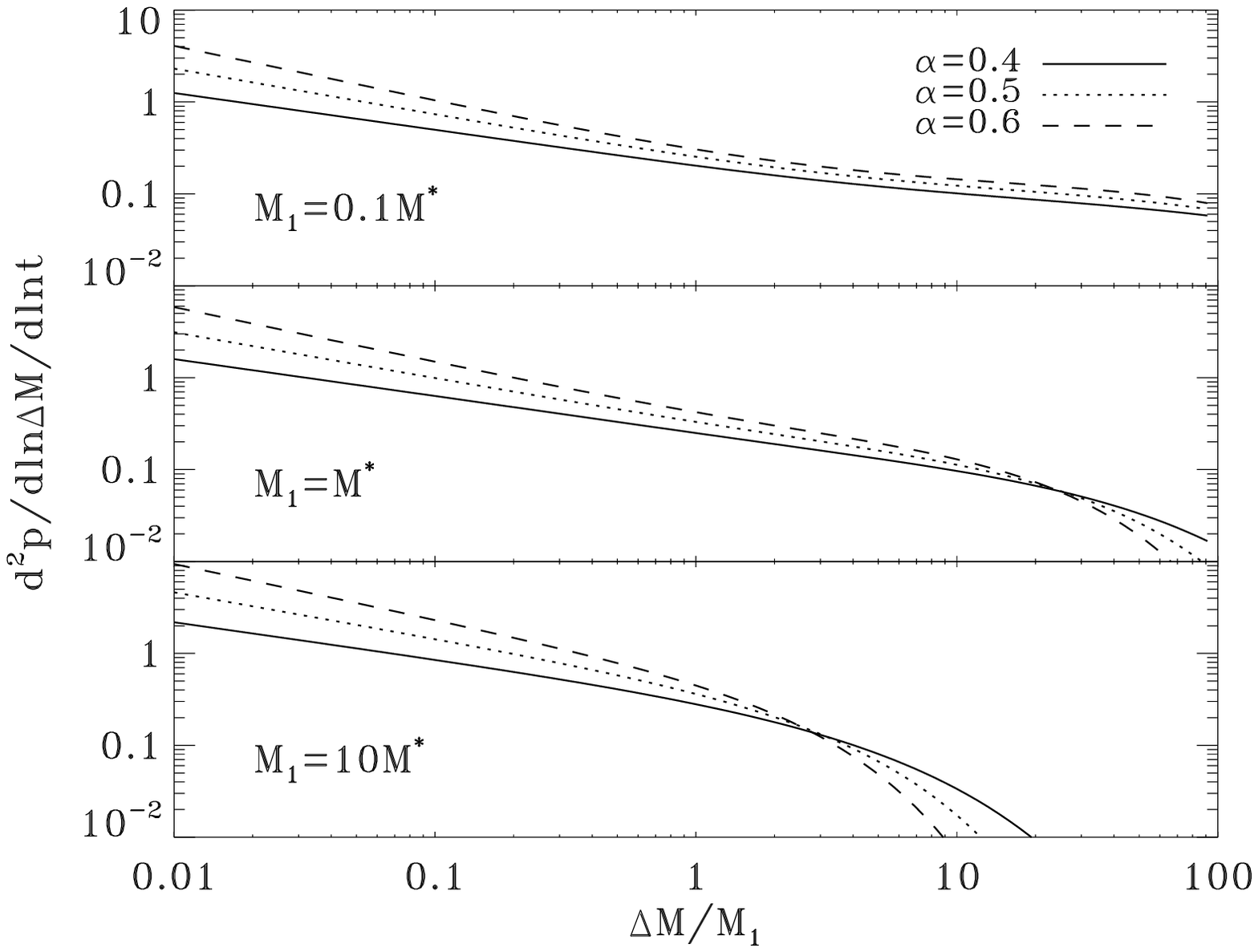}\includegraphics{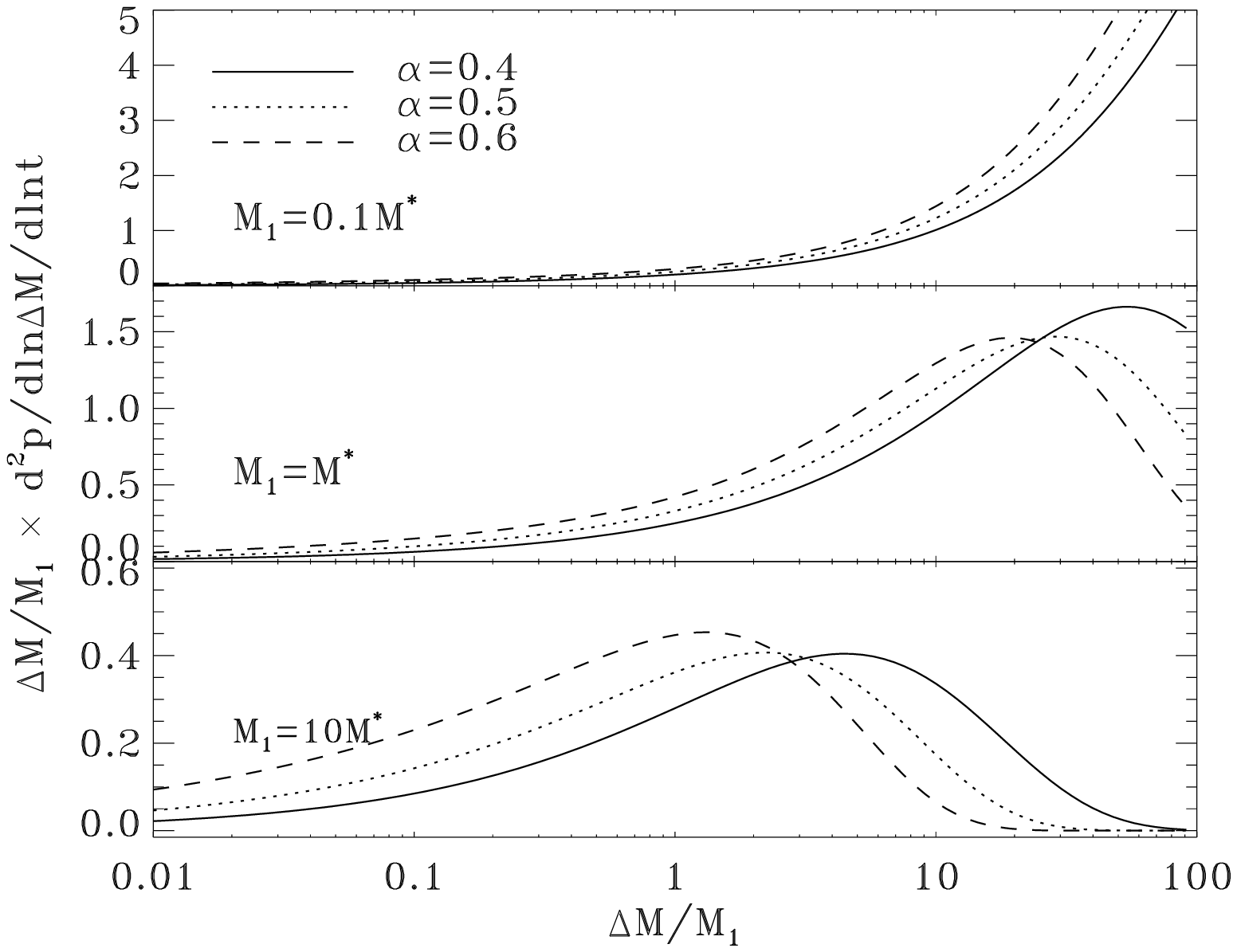}}
\caption{Halo merger rate (left panel) and accretion rate (right panel) computed
with Eq.~(\ref{eq:mr}). The accretion rate is simply
$\Delta M/M_1 \times d^2p/d\ln\Delta M/d\ln t$. Hereafter 
$M^*=10^{13} M_{\sun}$.}
\label{fig:mr}
\end{figure*}

Now we will
consider the merger rate function in line with \citet{LaceyCole1993}.
For a trajectory which has experienced a first up-crossing 
over $\omega_1$ at $S_1$, the conditional probability of having  
a first up-crossing over $\omega_2 (\omega_2 < \omega_1)$ at $S_2 (S_2 < S_1)$ 
in the interval $dS_2$ is given by the Bayes formula,
\begin{equation}
\begin{aligned}
f(S_2, \omega_2|& S_1,\omega_1)dS_2=\frac{f(S_1, \omega_1|S_2, \omega_2)dS_1 f(S_2, \omega_2)dS_2}
{f(S_1, \omega_1)dS_1} \\
= &\frac{2\alpha}{\sqrt{2\pi}}\frac{\omega_2(\omega_1-\omega_2)}{\omega_1}\left[ 
\frac{S_1}{S_2(S_1-S_2)} \right]^{\alpha+1} \\
\times & \exp \left[ - \frac{1}{2}\left(\frac{(\omega_1-\omega_2)^2}{(S_1-S_2)^{2\alpha}} +
\frac{\omega_2^2}{ S_2^{2\alpha}}
-\frac{\omega_1^2}{S_1^{2\alpha}}\right)\right] \ ,\\
(\omega_1>\omega_2\ , & \ S_1>S_2)\ .
\end{aligned}
\end{equation}
This function is interpreted as the probability that a halo of mass
$M_1$ at time $t$ will merge to build a halo of mass between $M_2$ and
$M_2+dM_2$ at time $t_2>t_1$. The mean transition rate is obtained from it
by setting $t_2\rightarrow t_1$ (equivalently $\omega_2\rightarrow \omega_1=\omega$),
\begin{equation}
\begin{aligned}
\frac{d^2 p(S_1\rightarrow S_2|\omega)}{dS_2 d\omega} & dS_2 d\omega=
\frac{2\alpha}{\sqrt{2\pi}} 
\left[ \frac{S_1}{S_2(S_1-S_2)} \right]^{\alpha+1} \\
\times & \exp\left[ -\frac{\omega^2}{2} 
\frac{S_1^{2\alpha} - S_2^{2\alpha}}{S_1^{2\alpha}S_2^{2\alpha}} \right]
dS_2 d\omega \ .
\end{aligned}
\end{equation}
Therefore the merger rate, i.e. the probability that a halo
of mass $M_1$ accretes a clump of mass $\Delta M=M_2-M_1$ within 
time $d\ln t$ (corresponding to $d\omega$), is
\begin{equation}
\begin{aligned}
d^2p&( M_1\rightarrow M_2|t)/d\ln \Delta M d\ln t \\
&= 2 \sigma(M_2) \Delta M \left| \frac{d\sigma_2}{dM_2} \right| 
\left| \frac{d\omega}{d\ln t} \right| 
\frac{d^2p(S_1\rightarrow S_2|\omega)}{dS_2d\omega}\\
&=2\alpha\sqrt{\frac{2}{\pi}}\frac{\Delta M}{M_2}
\left| \frac{d\ln\delta_c}{d\ln t} \right| 
\left| \frac{d\ln \sigma_2}{d\ln M_2}  \right| 
\frac{\delta_c(t)}{\sigma_2^{2\alpha}} \\
&\times \left( \frac{\sigma_1^2}{\sigma_1^2-\sigma_2^2} \right)^{\alpha+1}
\exp\left[ -\frac{\delta_c^2}{2} 
\left( \frac{1}{\sigma_2^{2\alpha}}- \frac{1}{\sigma_1^{2\alpha}} \right)
\right] \ .
\end{aligned}
\label{eq:mr}
\end{equation}

In Figure~\ref{fig:mr} the halo merger rates of different masses as predicted by
Eq.~(\ref{eq:mr}) are plotted for comparison. Three models are presented: 
an anti-persistent FBM of $\alpha=0.4$, the standard normal Brownian motion
of $\alpha=0.5$ and a persistent FBM of $\alpha=0.6$. 

For halo progenitors of the typical mass scale $M_1=M^*$, the merger rate and 
halo accretion rate increases with increasing $\alpha$ at small $\Delta M/M_1$,
but at very large mass ratios ($\Delta M/M_1>25$) the case is reversed. 

This is also true for more massive progenitor masses ($M_1=10M^*$), but here
the transition point in the mass ratio is smaller: $\Delta M/M_1 =3$. 

For the less massive progenitors, $M_1=0.1 M^*$, the transition point is very 
high, that in the whole plotted range (up to $\Delta M/M_1=100$), the models 
with greater Hurst exponent always have greater merger and accretion rate.

\subsection{Halo formation time distribution}

\begin{figure*}
\resizebox{\hsize}{!}{
\includegraphics{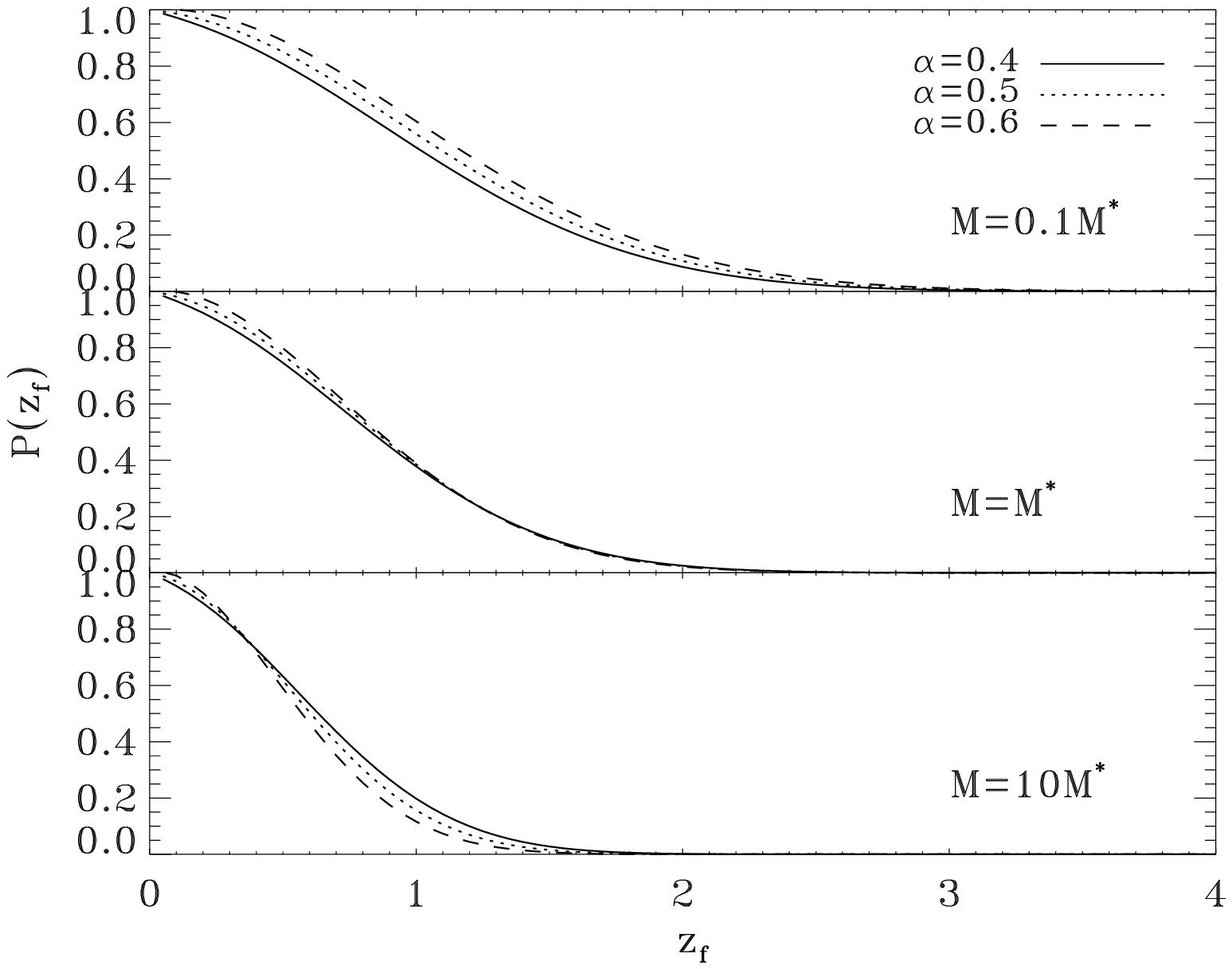}
\includegraphics{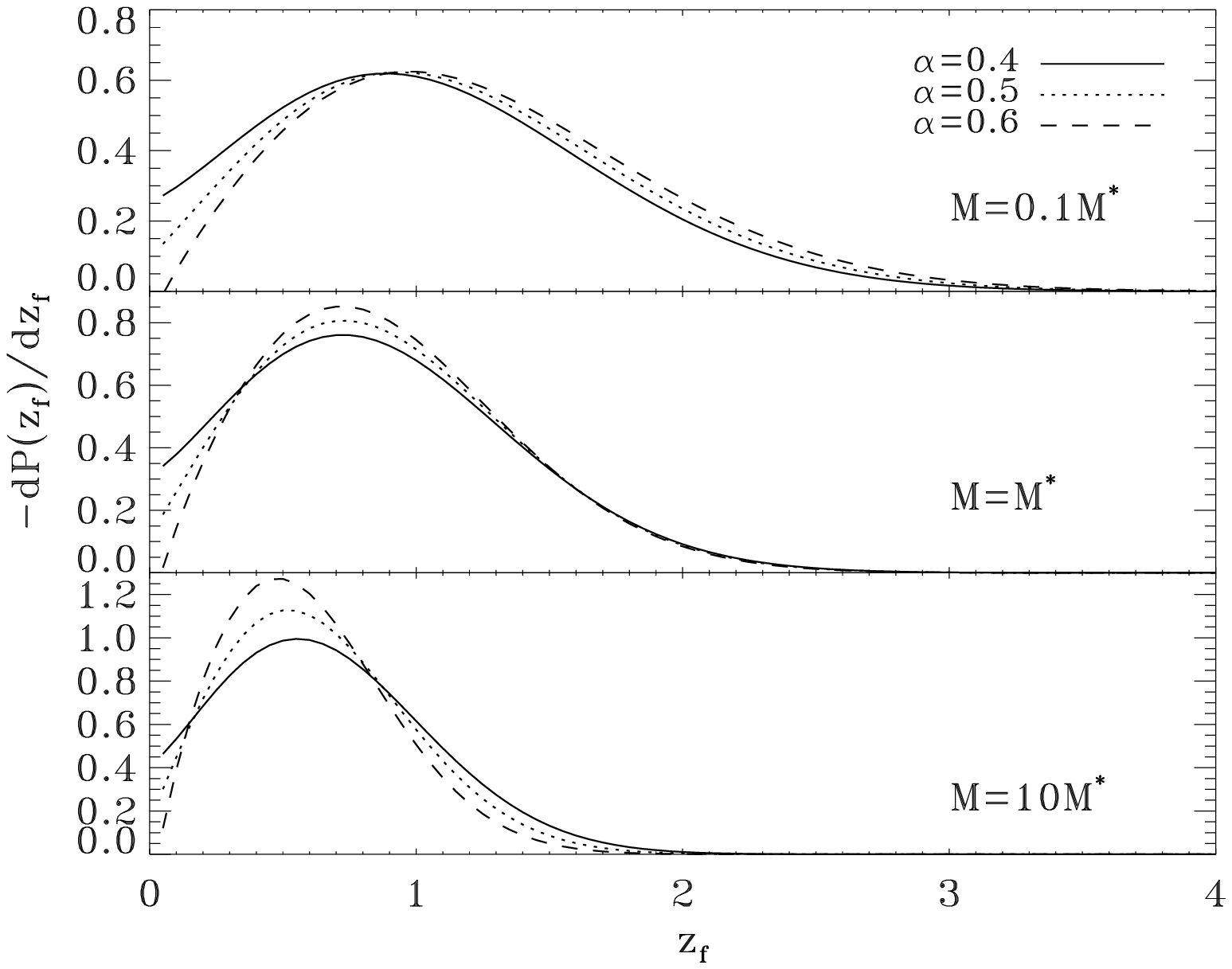}}
\caption{Halo formation time distributions for halos at $z=0$
with mass of $0.1$, 1 and $10M^{\star}$, which are predicted by the modified
excursion set theory of different $\alpha$ assuming spherical collapse. The 
left panel is the cumulative distribution.} 
\label{fig:ft}
\end{figure*}

The formation time (redshift) of a halo is the time (redshift) when its major
progenitor contains half of the halo mass. According to 
the counter argument of LC93, the cumulative distribution of halo formation
time, i. e. the probability that a halo of mass $M_0$ at redshift
$z_0$ is formed at redshift larger than $z_f$, can be acquired
by
\begin{equation}
P(>z_f)=\int_{S_0}^{S_h}\frac{M_0}{M(S_1)}
f\left(S_1,\delta_c(z_f)|S_0,\delta_c(z_0)\right)d S_1
\end{equation}
where $S_0=S(M_0)$ and $S_h=S(M_0/2)$. The halo formation time
distribution is simply given by the differentiation 
$-dP(z_f)/dz_f$.

Figure~\ref{fig:ft} shows the predictions of the halo formation
redshift distribution for difference masses. 
In general smaller halos have more extended formation time distribution 
than that of larger halos, 
in agreement with the results of 
\citet{LinEtal2003}. For $\alpha>1/2$, the halo
formation redshift distribution is more concentrated than the
$\alpha=1/2$ case, and for $\alpha<1/2$ the halo formation
redshift distribution is more extended. This seems to be in
accordance with our finding that the merger rate is greater for
larger $\alpha$. 

As can be seen, the detailed
effects of incorporating correlation between merging steps are
intricate, varying with the halo mass. Impact on the halo formation time
distribution is small for those halos with masses around $M^*$, 
and it becomes apparent only when halo 
mass deviates significantly from $M^*$.  The impact is 
also different for halos with 
small masses and large masses. Compared with the results given by
the standard excursion set theory ($\alpha=1/2$), 
\begin{enumerate}
\item $M \ll M^*$: 
negative correlation ($\alpha<1/2$) shifts the halo formation time distribution 
curve to the side of smaller $z_f$, which means that the younger halos are more 
abundant and the older ones are less abundant; whilst for
positive correlation ($\alpha>1/2$) the older ones are more abundant;
\item $M > M^*$: the impact of the correlation is opposite to the case of small
halo mass, negative correlation ($\alpha<1/2$) boosts more halos 
to form at an earlier time while positive correlation ($\alpha>1/2$)
induces more halos to form at an later time.
\end{enumerate}

\section{Ellipsoidal collapse}

\subsection{Moving barriers}
The spherical collapse model is perhaps accurate at high redshift e.g. 
the re-ionization era. However,  as shown by simulations, at low
redshift the collapse of a clump of mass is ellipsoidal.
For the excursion set theory, the general collapse condition for 
halo formation is not a constant $\delta_c$ any more, but a
moving barrier ${\mathcal B}(S)$.

Imposing a moving barrier on the random walk of 
$(\widetilde{\delta},\Delta S)$ described by Eq.~(\ref{eq:dfFBM})
is equivalent to the case of placing a constant barrier boundary condition to
the diffusion equation with an extra drifting term \citep{Zentner2007}. In
our case of FBM, the Fokker-Planck equation turns out to be
\begin{equation}
\frac{\partial Q_\alpha}{\partial \Delta S}=
\alpha {\Delta S}^{2\alpha-1} \frac{\partial^2 Q_\alpha}{ \partial \widetilde{\delta}^2}
+\frac{\partial \Delta{\mathcal B}}{\partial \Delta S}
\frac{\partial Q_\alpha}{\partial \widetilde{\delta}}
\label{eq:dfECFBM}
\end{equation}
where $\Delta {\mathcal B}={\mathcal B}(S_1)-{\mathcal B}(S_0)$. Unfortunately
this equation can be solved analytically only for a few very 
special cases (see Appendix B).

\subsection{Conditional mass function: the Sheth-Tormen approximation}

\begin{figure*}
\resizebox{\hsize}{!}{\includegraphics{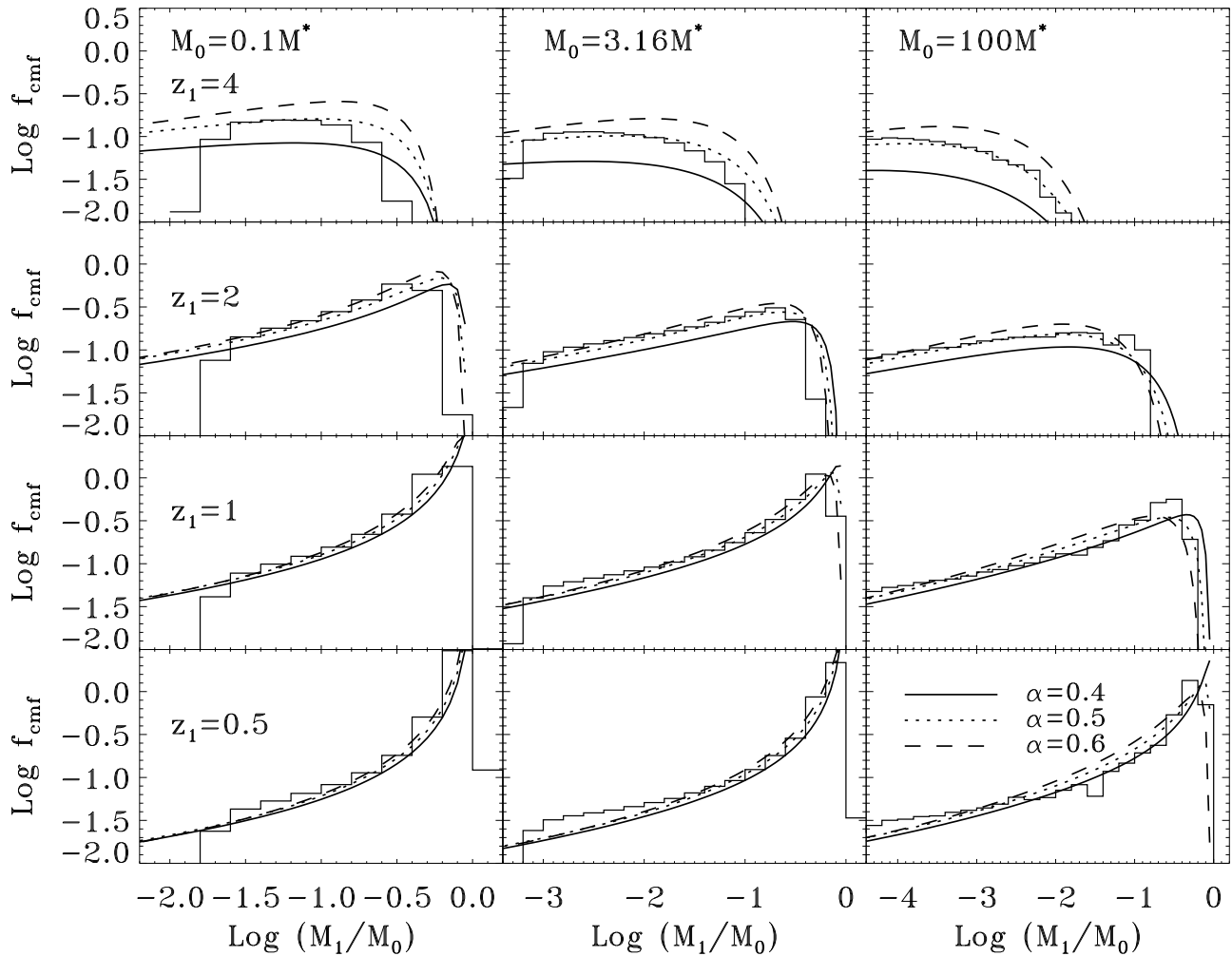}}
\caption{Conditional mass functions of progenitor halos at redshift 
$z_1=$0.5, 1, 2, 4 (identical for each row) for three different halo masses
$M_0$ (each column has the same mass). Histograms are the results
from the Millennium Simulation given by 
\citet{ColeEtal2008}, while lines are predictions by Eq.~(\ref{eq:ecCMF})
of three different values of $\alpha$ as indicated.} 
\label{fig:ecCMF}
\end{figure*}

Although we do not have analytical solution of Eq.~(\ref{eq:dfECFBM}),
it could be solved with the numerical method of \citet{ZhangHui2006}. However,
this is inconvenient for general explorations. Here
we borrow the pragmatic approach of \citet{ShethTormen2002},
the conditional mass function is then approximated by
\begin{equation}
\begin{aligned}
f(S_1-S_0)dS_1& =dS_1 \frac{2\alpha}{\sqrt{2\pi}}
\frac{\left|T(S_1, z_1|S_0,z_0)\right|}{(S_1-S_0)^{\alpha+1}}\\
\times & 
\exp\left\{ - \frac{\left[ {\mathcal B}(S_1, z_1)-{\mathcal B}(S_0,z_0) \right]^2}
{2 (S_1-S_0)^{2\alpha}} \right\}\ ,
\end{aligned}
\label{eq:ecCMF}
\end{equation}
in which
\begin{equation}
T(S_1|S_0)=\sum_{n=0}^5 \frac{(S_0-S_1)^n}{n!}
\frac{\partial^n \left[{\mathcal B}(S_1)-{\mathcal B}(S_0) \right]}{{\partial S_1}^n}
\label{eq:ecT}
\end{equation}
and 
\begin{equation}
{\mathcal B}(S, z)=\sqrt{q} \delta_c(z) \left[ 1+ b \left( \frac{S}{q \delta_c^2(z)}
\right)^\gamma \right]
\end{equation}
with $q=0.707$, $b=0.485$ and $\gamma=0.615$ 
\citep{ShethMoTormen2001, ShethTormen2002}.

Inspired by the results given in Appendix B, 
it is probably more appropriate to use
\begin{equation}
T=\sum_n \frac{ (-\widetilde{S})^n}{n!} 
\frac{\partial^n \Delta {\mathcal B}}{{\partial \widetilde{S}}^n}
\end{equation}
with $\widetilde{S}=(S_1-S_0)^{2\alpha}$ in Eq.~\ref{eq:ecCMF}. 
Testing the two approximation methods numerically, we found that
the actual difference between the two is in fact very small. 
So we opted to use the numerically simpler
Eq.~(\ref{eq:ecT}).

In figure~\ref{fig:ecCMF} we compare our prediction on 
the conditional mass function with that measured from the
Millennium Simulation \citep{ColeEtal2008} \footnote{The data is publicly
available at {\em http://star-www.dur.ac.uk/$\sim$cole/merger\_trees}} at
$z_0=0, z_1=0.5, 1, 2, 4$, for halos of masses $M_0$=0.1, 3.16, 100$M^*$. 
It is indeed very interesting to notice that the performance of
the original Sheth-Tormen approach (Eq.~\ref{eq:ecCMF} of $\alpha=1/2$) is
fairly good though not as remarkable as the fitting function
of \citet{ColeEtal2008}. It is surprising that
\citet{ColeEtal2008} extrapolated the Sheth-Tormen mass function for 
the conditional mass function in ellipsodal collapse model, in spite of
the one actually proposed in \citet{ShethTormen2002}.

From Figure~\ref{fig:ecCMF} we can see that the effects of correlation
between merging steps gradually decrease with time, with halo formation 
becoming sensitive to the correlation at high redshift. 
Except in cases of high
redshifts and of extremely high halo mass, it appears that
positive correlation ($\alpha>1/2$) is preferred by the simulation 
in $\Lambda$CDM universe. Nevertheless, one must be cautious to
this result, as the measurement presented is rather
crude and lacks error bars.  Its accuracy is not sufficient to justify with 
confidence whether the correlation between merging steps is positive, 
negative or zero.

\subsection{Halo formation time distribution}
\begin{figure*}
\resizebox{\hsize}{!}{
\includegraphics{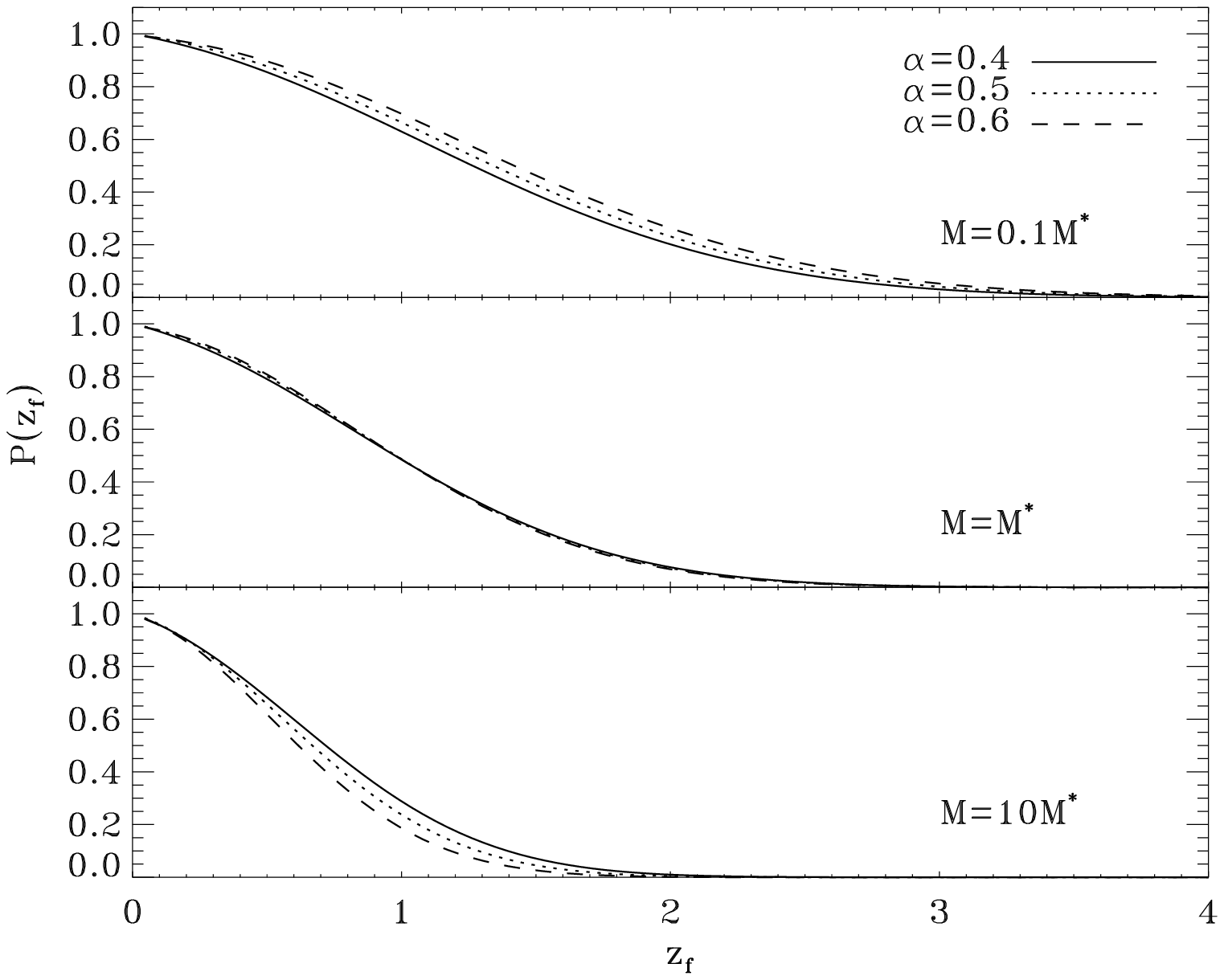}
\includegraphics{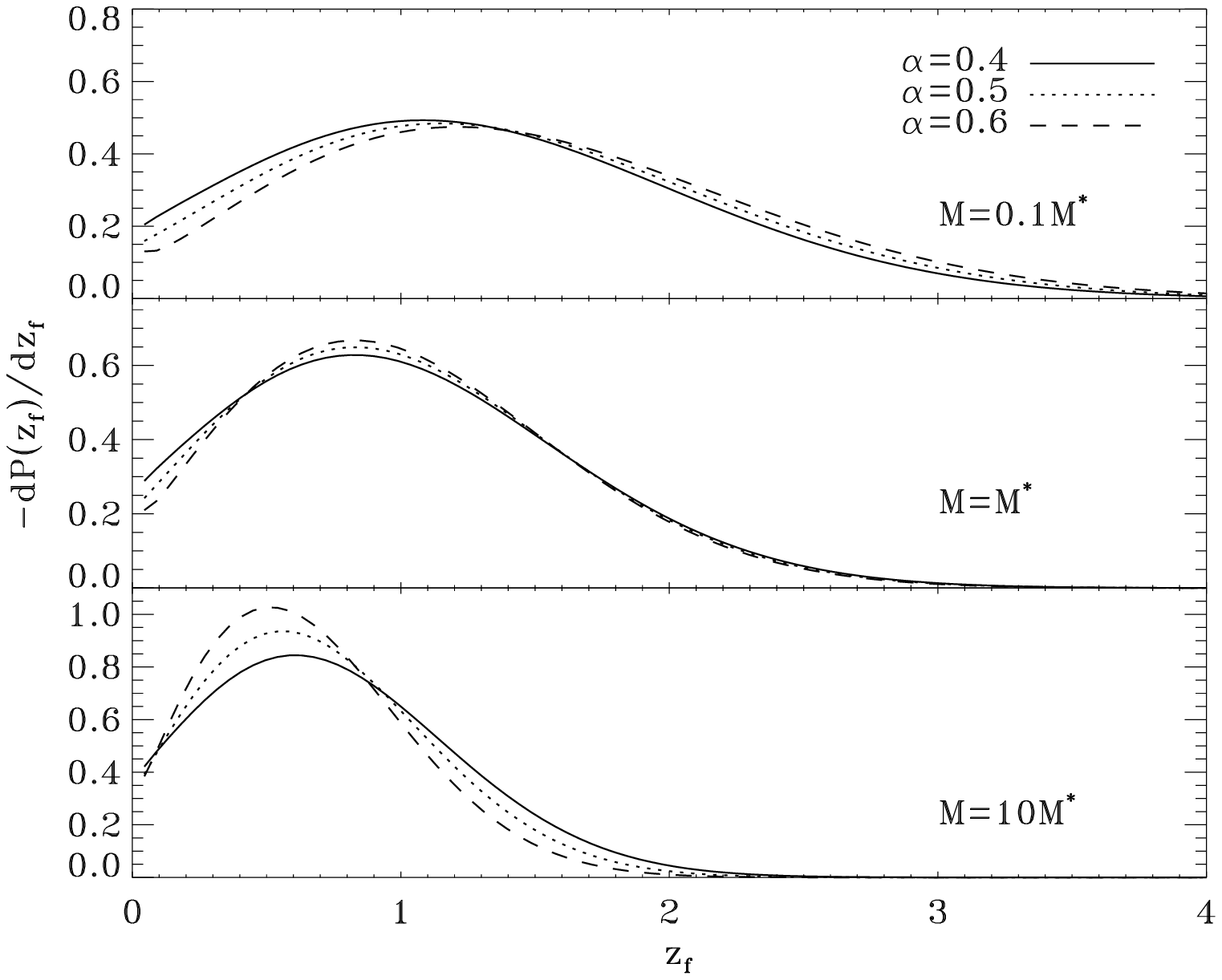}}
\caption{Halo formation time distributions for halos of masses 
0.1, 1 and 10$M^{\star}$ at $z=0$, which are predicted by the modified
excursion set theory of different $\alpha$ in ellipsoidal collapse model. 
The left panel is the cumulative distribution.} 
\label{fig:ecft}
\end{figure*}

The halo formation time distribution is recalculated with the modified
excursion set theory in ellipsoidal collapse model and the results 
displayed in Figure~\ref{fig:ecft}. Although the halo formation time distributions
are modulated significantly by the non-constant barrier, 
its dependence on $\alpha$ does not change, and demonstrates similar trends as
the spherical models predicts:
positive correlation results in more old small mass halos and more young
large mass halos, while the effects of the 
negative correlation are to
the contrary.

\citet{GiocoliEtal2007} calculated the cumulative distribution of halo formations
time and compared with GIF2 simulation. They found that the median formation
redshift of halos of mass less than $M^*$ extracted from the simulation
is larger than that predicted by the Sheth-Tormen formula, even after some
numerical improvement. Combined with our  discovery of effects of correlation, 
it seems that the discrepancy can be explained by the existence
of a positive correlation between adjacent merging steps.

\section{Conclusions and discussion}
In this work the FBM is formally incorporated in the excursion 
set theory, to account for possible correlations between merging steps 
of halo formation. Such modification
is minimal but provides a unified theoretical frame to investigate the
effects of correlation. More specifically, the correlation depends on the
Hurst exponent $\alpha$, for $\alpha<1/2$, the correlation
between adjacent merging steps is negative, for $\alpha>1/2$ the 
correlation is positive, while $\alpha=1/2$ is reduced to the standard case
of null correlation.

We calculated the conditional mass functions in models 
of the spherical collapse and the ellipsoidal collapse. Different
collapsing conditions do not change the relative effects of
correlations. It is revealed that for a typical progenitor of $M^*$, 
a positive correlation will boost 
the merger rate  when the mass to accrete is less than 
$\sim 25M^*$, and negative correlation will significantly reduce
such a rate.
Accordingly we find that compared with the
standard excursion set theory without correlation,
positive correlation increases the aged population for small mass halos
but produces more young members for large mass halos; while the negative
correlation projects opposite effects to the halo formation. The same trend is 
also seen for other progenitor masses, although the transition mass ratio
is greater for small progenitor masses and smaller for greater progenitor 
masses.

The results of the modified excursion set theory in ellipsoidal
collapse model are checked with those measurements from simulations of
\citet{GiocoliEtal2007} and \citet{ColeEtal2008}. The comparison
indicates that there is a sign of weak positive 
correlation, although not yet of significant confidence, as it is 
limited by the accuracy and dynamic ranges of their analysis. 
The conclusion appears to be in conflict with the claim of 
\citet{Pan2007}, but note that in that work 
the mass function was calculated with the spherical 
collapse model, which is not valid at most redshifts of concern.
Besides, the main purpose of \citet{Pan2007} is to show that 
the effect of correlation is not trivial.

The accuracy of the Sheth-Tormen approximation to the ellipsoidal
collapse in the modified (FBM) excursion set theory has not been 
checked against rigorous numerical solutions. However,  
since the actual correlation is weak, 
i.e. $\alpha$ only deviates from $1/2$ slightly, the
approximation suffices for general qualitative discussion. 
Calibration and subsequent improvement of the approximation
with numerical computation is of course needed
for more precise analytical modeling and implementation
in semi-analytical models of galaxy formation.

In this paper we explored the effects of correlations between 
mergers on the global halo formation process.  It is of 
interests to actually check if such a correlation will 
lead to the assembly bias discovered by \citet{GaoEtal2005}, and whether it
can explain the dependence of halo formation on the large scale environment.

A by product of our work is the discovery of a systematic effect
in the mass function due to the finite volume of simulation. It explains
why applying a high mass cut-off to the mass function can improve
the halo model prediction on three point correlation function. A better
approach in the halo model calculation of three point function 
is to shift the theoretical mass function
according to the formulae given in Appendix A, this may lead to better
agreement between simulation and analytical results.

\section*{Acknowledgment}
The authors acknowledge stimulating discussions with
Pengjie Zhang, Jun Zhang, Weipeng Lin and Longlong Feng.
JP is supported by the China Ministry of Science \& Technology 
through 973 grant of No. 2007CB815402 and the NSFC through
grants of Nos. 10643002, 10633040. YGW and XLC
are supported by NSFC via grants 1052314, 10533010, the CAS
under grant KJCX3-SYW-N2, and the Ministry of
Science \& Technology via 973 grant of 2007CB815401.
LT acknowledges the financial support of the Leverhulme Trust (UK).


\begin{thebibliography}{}

\bibitem[\protect\citeauthoryear{{Bond}, {Cole}, {Efstathiou} \&
  {Kaiser}}{{Bond} et~al.}{1991}]{BondEtal1991}
{Bond} J.~R.,  {Cole} S.,  {Efstathiou} G.,    {Kaiser} N.,  1991, \apj, 379,
  440

\bibitem[\protect\citeauthoryear{{Cole}, {Helly}, {Frenk} \&
  {Parkinson}}{{Cole} et~al.}{2008}]{ColeEtal2008}
{Cole} S.,  {Helly} J.,  {Frenk} C.~S.,    {Parkinson} H.,  2008, \mnras, 383,
  546

\bibitem[\protect\citeauthoryear{{Cuesta}, {Prada}, {Klypin} \&
  {Moles}}{{Cuesta} et~al.}{2007}]{CuestaEtal2007}
{Cuesta} A.~J.,  {Prada} F.,  {Klypin} A.,    {Moles} M.,  2007,
  astro-ph/0710.5520

\bibitem[\protect\citeauthoryear{{Feder}}{{Feder}}{1988}]{Feder1988}
{Feder} J.,  1988, {Fractals}.
Plenum Press, New York

\bibitem[\protect\citeauthoryear{{Fosalba}, {Pan} \& {Szapudi}}{{Fosalba}
  et~al.}{2005}]{FosalbaPanSzapudi2005}
{Fosalba} P.,  {Pan} J.,    {Szapudi} I.,  2005, \apj, 632, 29

\bibitem[\protect\citeauthoryear{{Gao}, {Springel} \& {White}}{{Gao}
  et~al.}{2005}]{GaoEtal2005}
{Gao} L.,  {Springel} V.,    {White} S.~D.~M.,  2005, \mnras, 363, L66

\bibitem[\protect\citeauthoryear{{Giocoli}, {Moreno}, {Sheth} \&
  {Tormen}}{{Giocoli} et~al.}{2007}]{GiocoliEtal2007}
{Giocoli} C.,  {Moreno} J.,  {Sheth} R.~K.,    {Tormen} G.,  2007, \mnras, 376,
  977

\bibitem[\protect\citeauthoryear{{Lacey} \& {Cole}}{{Lacey} \&
  {Cole}}{1993}]{LaceyCole1993}
{Lacey} C.,  {Cole} S.,  1993, \mnras, 262, 627

\bibitem[\protect\citeauthoryear{{Lin}, {Jing} \& {Lin}}{{Lin}
  et~al.}{2003}]{LinEtal2003}
{Lin} W.~P.,  {Jing} Y.~P.,    {Lin} L.,  2003, \mnras, 344, 1327

\bibitem[\protect\citeauthoryear{{Luki{\'c}}, {Heitmann}, {Habib}, {Bashinsky}
  \& {Ricker}}{{Luki{\'c}} et~al.}{2007}]{LukicEtal2007}
{Luki{\'c}} Z.,  {Heitmann} K.,  {Habib} S.,  {Bashinsky} S.,    {Ricker}
  P.~M.,  2007, \apj, 671, 1160

\bibitem[\protect\citeauthoryear{Lutz}{Lutz}{2001}]{Lutz2001}
Lutz E.,  2001, Phys. Rev. E, 64, 051106

\bibitem[\protect\citeauthoryear{{Pan}}{{Pan}}{2007}]{Pan2007}
{Pan} J.,  2007, \mnras, 374, L6

\bibitem[\protect\citeauthoryear{{Reed}, {Bower}, {Frenk}, {Jenkins} \&
  {Theuns}}{{Reed} et~al.}{2007}]{ReedEtal2007}
{Reed} D.~S.,  {Bower} R.,  {Frenk} C.~S.,  {Jenkins} A.,    {Theuns} T.,
  2007, \mnras, 374, 2

\bibitem[\protect\citeauthoryear{{Sheth}}{{Sheth}}{1998}]{Sheth1998}
{Sheth} R.~K.,  1998, \mnras, 300, 1057

\bibitem[\protect\citeauthoryear{{Sheth}, {Mo} \& {Tormen}}{{Sheth}
  et~al.}{2001}]{ShethMoTormen2001}
{Sheth} R.~K.,  {Mo} H.~J.,    {Tormen} G.,  2001, \mnras, 323, 1

\bibitem[\protect\citeauthoryear{{Sheth} \& {Tormen}}{{Sheth} \&
  {Tormen}}{2002}]{ShethTormen2002}
{Sheth} R.~K.,  {Tormen} G.,  2002, \mnras, 329, 61

\bibitem[\protect\citeauthoryear{{Wang}, {Yang}, {Mo}, {van den Bosch} \&
  {Chu}}{{Wang} et~al.}{2004}]{WangEtal2004}
{Wang} Y.,  {Yang} X.,  {Mo} H.~J.,  {van den Bosch} F.~C.,    {Chu} Y.,  2004,
  \mnras, 353, 287

\bibitem[\protect\citeauthoryear{{Warren}, {Abazajian}, {Holz} \&
  {Teodoro}}{{Warren} et~al.}{2006}]{WarrenEtal2006}
{Warren} M.~S.,  {Abazajian} K.,  {Holz} D.~E.,    {Teodoro} L.,  2006, \apj,
  646, 881

\bibitem[\protect\citeauthoryear{{Zentner}}{{Zentner}}{2007}]{Zentner2007}
{Zentner} A.~R.,  2007, International Journal of Modern Physics D, 16, 763

\bibitem[\protect\citeauthoryear{{Zhang} \& {Hui}}{{Zhang} \&
  {Hui}}{2006}]{ZhangHui2006}
{Zhang} J.,  {Hui} L.,  2006, \apj, 641, 641

\end{thebibliography}


\appendix
\section{Finite volume effects in small-box simulations}

N-body simulations are always performed in cubic boxes of finite size 
$L_{sim}$. By definition, the density contrast of whole the simulation box is 
zero, but the variance is not. Henceforth the measured mass function 
of simulation is 
conditional, determined by $f(S_1=S, \omega_1=\delta_c| S_0(L_{sim})\neq 0, \omega_0=0)$. 
From Eq.~(\ref{eq:sccmf}) the explicit halo mass function of a simulation is then
\begin{equation}
\begin{aligned}
f(\sigma, L_{sim})& d\ln \sigma = \frac{4\alpha}{\sqrt{2\pi}} 
\frac{\delta_c}{(\sigma^2-\sigma_0^2)^\alpha}
\frac{\sigma^2}{\sigma^2-\sigma_0^2} \\
& \times \exp\left[ -\frac{\delta_c^2}{2(\sigma^2-\sigma_0^2)^{2\alpha}}\right] 
d\ln \sigma\ ,
\end{aligned}
\label{eq:simmf}
\end{equation}
where $$S=\sigma^2=(2\pi)^{-3}\int_{2\pi/L_{sim}}^\infty P(k)\widetilde{W}4\pi k^2dk,$$ and 
$$S_0=\sigma_0^2(L_{sim})=(2\pi)^{-3}\int_{2\pi/L_{sim}}^\infty P(k)
\widetilde{W}(k, L_{sim})4\pi k^2dk.$$ 
The infrared cutoff in the integration for $\sigma^2$ accounts for the 
deficiency of large scale power in finite box, which has already 
been identified as a systematic effect \citep[e.g.][]{ReedEtal2007}. 

\begin{figure}
\resizebox{\hsize}{!}{\includegraphics{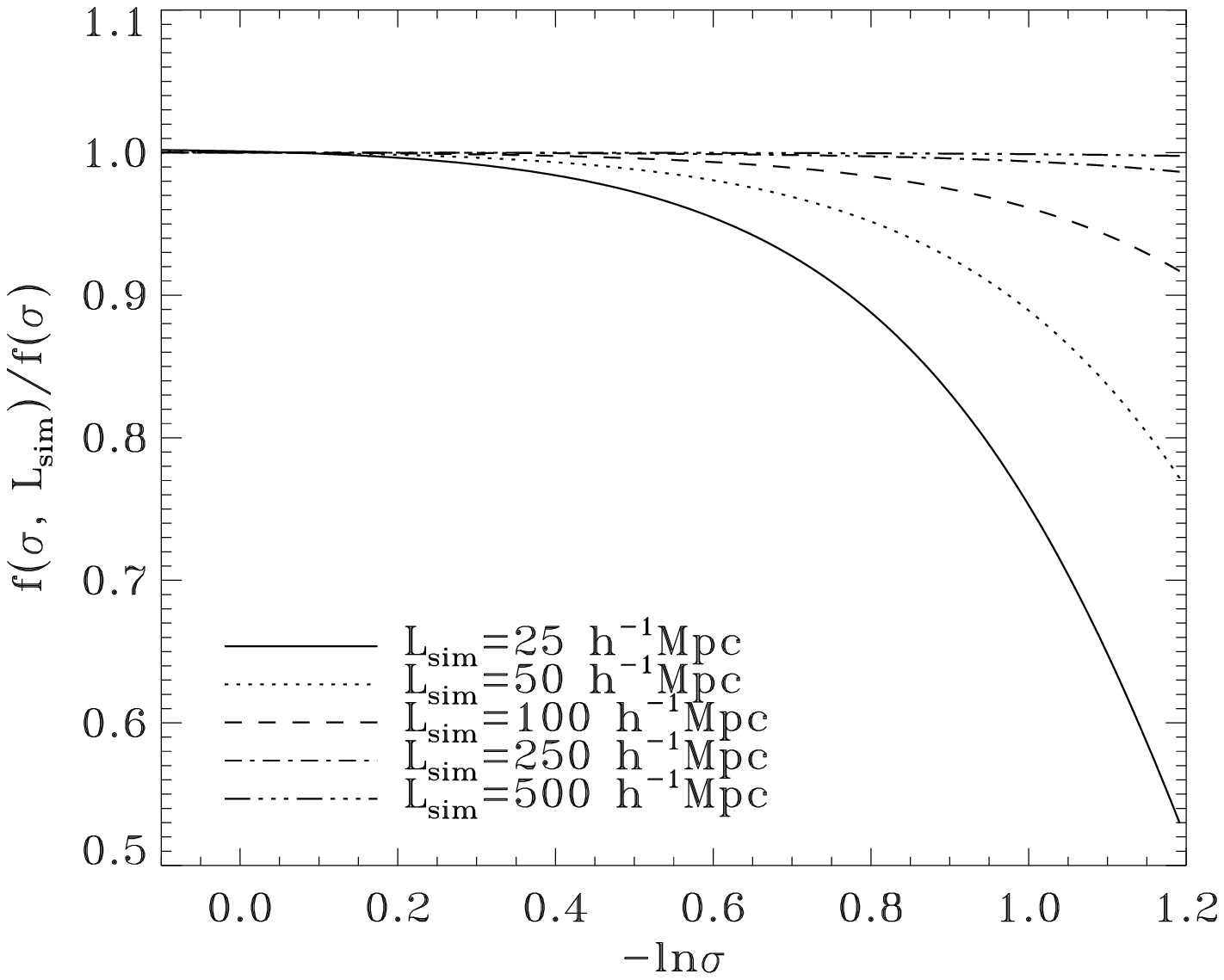}}
\caption{Deviation of mass functions in finite-sized simulations 
(Eq.~\ref{eq:simmf}) to Eq.~(\ref{eq:scmf}). For the fiducial  
model $\alpha=0.436$. 
$L_{sim}$ is the size of the cubic box for simulation. $\sigma_0$ is 
calculated with infrared cutoff in integration (see text), and a 
spherical top-hat window function in real space of radius 
$L_{sim}/(4\pi/3)^{1/3}$. Deviation is negligible in 
regime $-\ln \sigma <0.2$.}
\label{fig:simmf}
\end{figure}

The difference between Eq.~(\ref{eq:simmf}) and Eq.~(\ref{eq:scmf}) is shown in 
Fig.~\ref{fig:simmf} for different box sizes. The fiducial model has 
$\alpha=0.436$, by which the spherical model provides good approximation to 
the mass function of simulations in high mass region \citep{Pan2007}. We 
can see that the finite volume effect becomes apparent only when 
$-\ln \sigma >0.2$, i.e. in high mass regime, and the resulting
 fractional error 
quickly drops down to $<10\%$ for simulations of $L_{sim}>100h^{-1}$Mpc.

Nowadays it is a very common practice to use 
simulations with boxes as small as $< 50h^{-1}$Mpc to 
extend the dynamic range. The finite volume 
effect demonstrated in Fig.~\ref{fig:simmf} has to be taken into account 
for precision measurement. The necessary
correction is actually very simple: the true measured mass 
function $f(S)dS$ can be recovered 
by shifting the raw mass function $f(S|S_0)dS$ along $S$ by $S_0$. Note 
that there is a cosmic variance for $S_0$, for each individual 
realization of simulation one may need to calculate the individual 
$S_0$ from the realization \citep{ReedEtal2007, LukicEtal2007}.

\section{{\em pseudo-linear} barrier: 
${\mathcal B}=\delta_c+\beta S^{2\alpha}$}
Starting with Eq.~(\ref{eq:dfECFBM}), let $S_0=0$ and $S_1=S>0$, 
a variable substitution $\widetilde{S}=S^{2\alpha}$ yields
\begin{equation}
\frac{\partial Q_\alpha}{\partial \widetilde{S}}=
\frac{1}{2}\frac{\partial^2 Q_\alpha}{ \partial \widetilde{\delta}^2}
+\frac{1}{2\alpha S^{2\alpha-1}}\frac{\partial {\mathcal B}}{\partial S}
\frac{\partial Q_\alpha}{\partial \widetilde{\delta}}
\end{equation}
If the barrier is ${\mathcal B}=\delta_c+\beta S^{2\alpha}$, the above 
nonlinear Fokker-Planck equation is further simplified to
\begin{equation}
\frac{\partial Q_\alpha}{\partial \widetilde{S}}=
\frac{1}{2}\frac{\partial^2 Q_\alpha}{ \partial \widetilde{\delta}^2}
+\beta \frac{\partial Q_\alpha}{\partial \widetilde{\delta}}\ ,
\end{equation}
which is the well known problem of a linear
barrier upon the normal Brownian motion \citep{Sheth1998, Zentner2007}.

The probability of a trajectory that has the first barrier passage 
within $(\widetilde{S}, \widetilde{S}+d\widetilde{S})$ in the solution
is given by
\begin{equation}
f(\widetilde{S})d\widetilde{S}=\frac{\delta_c}{\sqrt{2\pi} {\widetilde{S}}^{3/2}}
\exp\left[ - \frac{(\beta \widetilde{S}+\delta_c)^2}{2 \widetilde{S}} \right]
d\widetilde{S}\ ,
\end{equation}
from which the universal mass function is easily obtained
\begin{equation}
f(\sigma)d\ln \sigma=\frac{4\alpha}{\sqrt{2\pi}}\frac{\delta_c}{\sigma^{2\alpha}}
\exp\left[ - \frac{(\delta_c+\beta \sigma^{4\alpha})^2}{2\sigma^{4\alpha}}\right]
d\ln \sigma\ .
\label{eq:linBmf}
\end{equation}

However, there is no analogous analytical expression for the 
conditional mass function under this type of barrier, as the
barrier does not have the symmetry that a true linear barrier possesses: 
for any pair of points at $S_0$ and $S_1>S_0$
\begin{equation}
{\mathcal B}_1-{\mathcal B}_0=\beta (S_1^{2\alpha}-S_0^{2\alpha})\neq 
\beta(S_1-S_0)^{2\alpha}\ .
\end{equation}
This is why we call it {\em pseudo-linear}.

\end{document}